\documentclass[prd,preprintnumbers,twocolumn,aps]{revtex4-1}
\usepackage{amsmath, bm, amssymb, epsfig, graphicx, subfigure, dcolumn}
\usepackage[usenames,dvipsnames]{color}
\usepackage[pagebackref=false, colorlinks=true]{hyperref}
\definecolor{redish}{rgb}{0.7,0.2,0.0} 
\definecolor{bluish}{rgb}{0.2,0.5,0.8}
\hypersetup{linkcolor=redish, citecolor=blue, filecolor=magenta, urlcolor=bluish}          

\newcommand{\be}{\begin{equation}}
\newcommand{\ee}{\end{equation}}
\newcommand{\ba}{\begin{eqnarray}}
\newcommand{\ea}{\end{eqnarray}}
\newcommand{\ban}{\begin{eqnarray*}}
\newcommand{\ean}{\end{eqnarray*}}
\newcommand{\bwt}{\begin{widetext}}
\newcommand{\ewt}{\end{widetext}}

\newcommand{\Rlm}{\ensuremath{R_{lm}}}
\newcommand{\Slm}{\ensuremath{S_{lm}}}
\newcommand{\tR}{\ensuremath{\tilde{R}_{lm}}}
\newcommand{\ts}{\ensuremath{\tilde{s}}}
\newcommand{\tlambda}{\ensuremath{\tilde{\lambda}}}
\newcommand{\tomega}{\ensuremath{\tilde{\omega}}}

\begin{document}

\title{On the stability of a superspinar}

\author{Ken-ichi Nakao$^1$}
\author{Pankaj S. Joshi$^2$}
\author{Jun-Qi Guo$^{2,3}$}
\author{Prashant Kocherlakota$^2$}
\author{Hideyuki Tagoshi$^4$}
\author{Tomohiro Harada$^5$}
\author{Mandar Patil$^6$}
\author{Andrzej Kr\'{o}lak$^6$} 

\affiliation{
${}^{1}$Department of Mathematics and Physics, Graduate School of Science, Osaka City University, 3-3-138 Sugimoto, Sumiyoshi, Osaka 558-8585, Japan\\
${}^{2}$Tata Institute of Fundamental Research, Homi Bhabha Road, Mumbai 400005, India\\
${}^{3}$School of Physics and Technology, University of Jinan, No. 336, West Road of Nan Xinzhuang, Jinan 250022, Shandong, China \\
${}^4$Institute for Cosmic Ray Research (ICRR), University of Tokyo, 5-1-5 Kashiwa-no-ha, Kashiwa, Chiba 277-8582, Japan\\
${}^5$Department of Physics, Rikkyo University, Toshima, Tokyo 171-8501, Japan\\
$^6$ Institute of Mathematics, Polish Academy of Sciences, \'{S}niadeckich 8, 00-656 Warsaw, Poland
}


\begin{abstract}
The superspinar proposed by Gimon and Ho$\check{\rm r}$ava is a rapidly 
rotating compact entity whose exterior is described by the over-spinning Kerr geometry. The compact entity itself is expected to be governed by superstringy effects, and in astrophysical scenarios it can give rise to interesting observable phenomena. Earlier it was suggested that the superspinar may not be stable but we point out here that this does not necessarily follow from earlier studies. We show, by analytically treating the Teukolsky equations by Detwiler's method, that in fact there are infinitely many boundary conditions that make the superspinar stable, 
and that the modes will decay in time. It follows that we need to know more on 
the physical nature of the superspinar in order to decide on its stability in 
physical reality.  
\end{abstract}

\preprint{OCU-PHYS-468 }
\preprint{AP-GR-140 }
\preprint{RUP-17-14}
\date{\today}
\maketitle

\section{Introduction}

The Kerr spacetime is an exact stationary solution of the vacuum Einstein 
equations and is characterized by two parameters, namely the gravitational mass $M$ and the so-called Kerr parameter $a$ which is the angular momentum divided by $M$. The solution describes a rotating black hole if $a^2 \leq M^2$, whereas it describes a naked singular spacetime if $a^2>M^2$, using the geometrized 
units ($G$ = $c$ = $1$). The Kerr black hole has been extensively studied in many scenarios which would be stable against linear perturbations. This may suggest the reliability of the weak version of the cosmic censorship hypothesis whose statement is, roughly speaking, the spacetime singularities formed from generic initial conditions are enclosed by event horizons. Also, many black-hole candidates, $i.e.$ objects described by the Kerr solution of $a^2<M^2$ have been found in our universe.  


Gimon and Ho$\check{\rm r}$ava pointed out an interesting fact that the supersymmetry does not imply the Kerr bound $a^2\leq M^2$, and hence if a very compact object of $a^2>M^2$ is found, it may be a signal of superstring theory. They named it the superspinar. The naked singularity will be made harmless by stringy effect. However, before the indication of Gimon and Ho$\check{\rm r}$ava, a study suggested the instability of the over-spinning Kerr spacetime $a^2>M^2$ \cite{Dotti2007}. After the superspinar possibility, few more studies were done on the stability of the over-spinning Kerr geometry by other researchers \cite{Dotti2008, Cardoso2008, Pani2010}, to suggest that the superspinar is unstable under various boundary conditions.  The variety of the boundary conditions is maximal in the study by Pani et al, which includes all the previous studies, and they concluded that the over-spinning Kerr geometry and thus the superspinar is unstable. However, it should be noted that in order to conclude so, we must show that the over-spinning Kerr geometry is unstable under all possible boundary conditions, since at present nobody knows the physical nature of the superspinar. From this standpoint, the numerical results obtained by Pani et al may not necessarily imply the instability of the superspinar. 

In this paper, in order to illustrate the stability problem of the superspinar, we analytically treat the linear perturbations in the near-extremal over-spinning Kerr spacetime by the manner devised by Detweiler \cite{Detweiler1980,Sasaki1990,Cardoso2004}. It turns out that under a variety of boundary conditions the modes decay in time and the superspinar is stable. 

This result may have intriguing implications on the existence and physics of very rapidly rotating compact objects in the Universe. 
It therefore follows from our results here that, at the very least, we need
a detailed study of physically allowed boundary conditions in order to decide on
the stability of superspinar or similar objects.   


\section{Teukolsky Equations}

The perturbations in the Kerr spacetime are governed by the Teukolsky equation \cite{Teukolsky1973}; Writing the master variable $\psi$ in the form $\psi=e^{-i\omega t+im\varphi}\Slm(\theta)\Rlm(r)$, the radial and angular 
Teukolsky equations are given by
\begin{align}
&\Delta^{-s}\frac{d}{dr}\left(\Delta^{s+1}\frac{d\Rlm}{dr}\right)  \nonumber\\
&+\left(\frac{K^2-2is(r-M)K}{\Delta}+4is\omega r-\lambda\right)\Rlm=0,\label{R-eq}\\
&\frac{1}{\sin\theta}\frac{d}{d\theta}\left(\sin\theta\frac{d\Slm}{d\theta}\right) + \biggl[\left(a\omega\cos\theta+s\right)^2  \nonumber\\
&-\left(\frac{m+s\cos\theta}{\sin\theta}\right)^2-s(s-1)+F\biggr]\Slm=0,\label{S-eq}
\end{align}
for the scalar ($|s|=0$), the electromagnetic ($|s|=1$) and gravitational ($|s|=2$) perturbations, 
where $F={}_sF^l_{m,\omega}$ with the integer $l$ larger than or equal to ${\rm max}(|m|,|s|)$ 
is the separation constant equivalent to the 
eigenvalue of Eq.~(\ref{S-eq}) with the boundary conditions of regularity at $\theta=0$ and $\pi$, 
$K:=(r^2+a^2)\omega-am$, $\lambda:=F+a^2\omega^2-2am\omega$, and $\Delta:=r^2-2Mr+a^2$. In the case of $a^2<M^2$, $r=r_\pm:=M\pm\sqrt{M^2-a^2}$ are real roots of $\Delta=0$; $r=r_+$ corresponds to the event horizon and $r=r_-$ is the location of the Cauchy horizon. In the extremal case, $a^2=M^2$, $r_+$ and $r_-$ agree with each other, and there is only one degenerate event horizon. In the case of $a^2>M^2$, i.e., the superspinar, there is no real root of $\Delta=0$, and correspondingly no event horizon exists.

In order to see whether the superspinar is stable, we investigate the angular frequencies of the quasi-normal modes, which are linear perturbations around the Kerr metric without incoming waves at infinity. Hence we focus on the component of the Weyl tensor denoted by $\psi_4$, which corresponds to outgoing gravitational waves and relates to the master variable through $\psi_4=(r-ia\cos\theta)^{-4}~\psi$ with $s=-2$. 

Hereafter, we follow Ref.~\cite{Cardoso2004}  so that it is easy to compare the superspinar case with the black-hole case. Instead of $\Rlm$, the following variable is introduced;
\begin{equation}
\Rlm=\Delta^{-s}\tR \exp\left(-i\int\frac{K}{\Delta}dr\right).
\end{equation}
Then, Eq.~(\ref{R-eq}) becomes
\begin{align}
&\Delta\frac{d^2\tR}{dr^2}-\left[2i\omega (r^2+a^2)-2(\ts+1)(r-M)-2iam\right] \nonumber\\ 
&\times\frac{d\tR}{dr} -\left[2(2\ts+1)i\omega r+\tilde{\lambda}\right]\tR=0, \label{tR-eq}
\end{align}
where, using $F=E-s(s+1)$, we have introduced 
$\ts:=-s$ and $\tlambda:=\lambda+2s=E+a^2\omega^2-2am\omega-\ts(\ts+1)$.

\section{Quasi-normal modes of near-extremal Kerr spacetime}

We consider a near-extremal Kerr spacetime and hence we write the Kerr parameter in the form $$ a=M(1-\epsilon), $$ assuming $0<|\epsilon|\ll1$. The spacetime contains a superspinar in the case of $\epsilon<0$, 
whereas there is a black hole in the case of $\epsilon> 0$. 

In the case of black hole, it is known that the quasi-normal mode (QNM) frequency $\omega$ approaches $m/2M$ for  $m=l$ in the limit of $\epsilon\rightarrow0_+$ \cite{Detweiler1980}. The numerical study in Ref.~\cite{Pani2010} has revealed that even in the superspinar case, the QNM frequency $\omega$ approaches $m/2M$ for $m=l$ modes in the limit $\epsilon\rightarrow0_-$. Hence, hereafter we focus on the modes of $m=l$ and assume 
\begin{equation}
M\omega-\frac{m}{2}={\cal O}\left(|\epsilon|^p\right), \label{assumption}
\end{equation}
where $p$ is a positive constant. 

We rewrite Eq.~(\ref{tR-eq}) in terms of the the dimensionless variables $y:=(r-M)/M$ and $\tomega:=M\omega$ as, 
\begin{align}
(y^2&-2\epsilon+\epsilon^2)\frac{d^2\tR}{dy^2} -\left[2i\tomega y^2+2(2i\tomega-\ts-1)y \right. \nonumber\\
& \left. +2i(2\tomega-m)(1-\epsilon) + 2i\tomega\epsilon^2\right]\frac{d\tR}{dy} \nonumber \\
& -\left[ 2(2\ts+1)i\tomega(y+1)+\tlambda\right]\tR=0.
\label{Basic-eq}
\end{align}

Before proceeding to our task, we briefly mention our strategy to obtain the QNM frequency for the black hole case. First, we obtain the approximate solutions of Eq.~(\ref{Basic-eq}) in the far zone defined as $y\gg {\rm max}\left[\sqrt{|\epsilon|},|\epsilon|^p\right]$ and the near zone defined as $y\ll1$, separately. 
Then, we choose appropriate integration constants so that these solutions agree with each other in the overlapping region, ${\rm max}\left[\sqrt{|\epsilon|},|\epsilon|^p\right] \ll y \ll1$. Finally, we impose the no-incoming wave condition on the far-zone solution at infinity and the regularity condition on the near-zone solution at the event horizon, for black holes. A similar procedure is followed for the superspinar in order to clarify the difference from the black hole case. 

In the far zone, the following equation approximates to Eq.~(\ref{Basic-eq}); 
\begin{align}
& y^2\frac{d^2\tR}{dy^2} -\left[2i\tomega y^2+2(2i\tomega-\ts-1)y\right]\frac{d\tR}{dy}\nonumber\\
& -\left[2(2\ts+1)i\tomega(y+1)+\tlambda\right]\tR=0. \nonumber
\end{align}
The solution of the above equation is written in terms of confluent hypergeometric functions $_1F_1(\alpha;\gamma;z)$; 
\begin{align}
\tR^{~\rm far}
&= A y^{-\ts-1/2+2i\tomega+i\delta} \nonumber\\
&\times{}_1F_1\left(\frac{1}{2}+\ts+2i\tomega+i\delta;1+2i\delta;2i\tomega y\right)\nonumber \\
&+B y^{-\ts-1/2+2i\tomega-i\delta}\nonumber\\
&\times{}_1F_1\left(\frac{1}{2}+\ts+2i\tomega-i\delta;1-2i\delta;2i\tomega y\right),
\label{solution-far}
\end{align}
where $A$ and $B$ are integration constants, and $\delta$ is a constant which 
can be found in Ref.~\cite{Cardoso2004}. 

For the near-zone analysis, we keep terms only of leading order in $\epsilon$ and introduce a new radial variable, $x:=y-\sqrt{2\epsilon}$. Then, Eq.~(\ref{Basic-eq}) approximates to,
\begin{align}
& x(x+\sigma)\frac{d^2\tR}{dx^2} -\Bigl[2\left(2i\tomega-\ts-1\right)x-\left(\ts+1\right)\sigma \nonumber \\
& +4i\tau\Bigr]\frac{d\tR}{dx} -\left[2(2\ts+1)i\tomega+\tlambda\right]\tR=0, \label{near-2}
\end{align}
where 
$$
\sigma :=2\sqrt{2\epsilon}~~~{\rm and}~~~
\tau :=(1+\sqrt{2\epsilon})\tomega-\frac{m}{2}.
$$
The solution of Eq.~(\ref{near-2}) is expressed by using  
Gauss's hypergeometric function ${}_2F_1(\alpha,\beta;\gamma;z)$ in the form 
\ba
\tR^{\ \rm near}&&=C~x^{-\ts+4i\tau/\sigma} {}_2F_1(1/2-2i\tomega+i\delta+4i\tau/\sigma, \nonumber \\
&& 1/2-2i\tomega-i\delta+4i\tau/\sigma; 1-\ts+4i\tau/\sigma;-x/\sigma) \nonumber \\
&& +D~ {}_2F_1(1/2+\ts-2i\tomega+i\delta,1/2+\ts-2i\tomega-i\delta;\nonumber \\
&&1+\ts-4i\tau/\sigma;-x/\sigma), \label{solution-near}
\ea
where $C$ and $D$ are integration constants. 

Both solutions (\ref{solution-far}) and (\ref{solution-near}) are valid in the over-lapping region.  
In the limit $y\rightarrow0$, the solution (\ref{solution-far}) behaves as
$
\tR\rightarrow Ay^{-\ts-1/2+2i\tomega+i\delta} +By^{-\ts-1/2+2i\tomega-i\delta}.
$
In the limit $y\rightarrow \infty$, the solution (\ref{solution-near}) behaves as
$
\tR\rightarrow {\cal A}y^{-\ts-1/2+2i\tomega+i\delta}+{\cal B} y^{-\ts-1/2+2i\tomega-i\delta},
$
where $\cal A$ and $\cal B$ are 
\begin{align}
&{\cal A}=\sigma^{1/2-2i\tomega-i\delta}\Gamma(2i\delta) \nonumber \\
&\times\Biggl[
\frac{C\sigma^{4i\tau/\sigma}\Gamma(1-\ts+4i\tau/\sigma)}
{\Gamma(1/2-\ts+2i\tomega+i\delta)\Gamma(1/2-2i\tomega+i\delta+4i\tau/\sigma)} \nonumber \\
&+\frac{D\sigma^{\ts}\Gamma(1+\ts-4i\tau/\sigma)}
{\Gamma(1/2+\ts-2i\tomega+i\delta)\Gamma(1/2+2i\tomega+i\delta-4i\tau/\sigma)} 
\Biggr], \nonumber\\
 \label{calA}\\
&{\cal B}={\cal A}|_{\delta\rightarrow-\delta}. \label{calB}
\end{align}
Thus, the matching condition is
\be
A={\cal A}~~~~{\rm and}~~~~B={\cal B}. \label{matching}
\ee
From the far-zone solution (\ref{solution-far}), for $y\rightarrow\infty$, we have
\be
\tR^{\ \rm far}\simeq Z_{\rm out}~ y^{-(1-4i\tomega)}e^{2i\tomega y} +Z_{\rm in}~ y^{-(2\ts+1)},\nonumber 
\ee
where
\begin{align}
Z_{\rm in}&=A\frac{(-2i\tomega)^{-1/2-\ts-2i\tomega-i\delta}\Gamma(1+2i\delta)}
{\Gamma(1/2-\ts-2i\tomega+i\delta)} \cr
&+B\frac{(-2i\tomega)^{-1/2-\ts-2i\tomega+i\delta}\Gamma(1-2i\delta)}{\Gamma(1/2-\ts-2i\tomega-i\delta)}, \cr
Z_{\rm out}&=Z_{\rm in}|_{\ts\rightarrow-\ts,\tomega\rightarrow-\tomega}. \nonumber
\end{align}
Thus, together with Eq.~(\ref{matching}), 
the no-incoming wave boundary condition, $Z_{\rm in}=0$, leads to 
\begin{equation}
{\cal A}\frac{(-2i\tomega)^{-i\delta}\Gamma(1+2i\delta)}{\Gamma(1/2-\ts-2i\tomega+i\delta)} 
+\left(\delta\rightarrow-\delta\right)=0
\label{QNM-condition}
\end{equation}
Here it is worthwhile to notice that, in the black hole case ($\epsilon>0$), the regular singular point $x=0$ of Eq.~(\ref{near-2}) corresponds to the location of the event horizon. Since we impose the regularity of the solution at the event horizon, the integration constant $C$ must vanish (note $\ts=2$). By contrast, in the superspinar case ($\epsilon<0$), the regular singular points $x=0$ and $x=-\sigma$ of Eq.~(\ref{near-2}) are equivalent to $y=\pm i\sqrt{2|\epsilon|}$. Hence, there is no regular singular point of Eq.~(\ref{near-2}) on the real axis of $y$, or equivalently, on the real axis of $r$. This is a distinctive feature of the superspinar from that of the black hole. The regularity requirement of the solution on the real axis of $y$ does not lead to any condition on the integration constants $C$ and $D$ in the superspinar case. However, in order to get the QNM frequency in the superspinar case, we need to fix $C$ and $D$. Thus, for example, we impose identical conditions for both the black hole ($\epsilon>0$) and the superspinar ($\epsilon<0$); 
\begin{equation}
C=0~~~{\rm and}~~~D=1. \label{BC-near-1}
\end{equation}
Substituting Eqs.~(\ref{calA}) and (\ref{calB}) with the condition (\ref{BC-near-1}) into  
Eq.~(\ref{QNM-condition}), we have
\begin{align}
-&\frac{\Gamma(2i\delta)\Gamma(1+2i\delta)}
{\Gamma(-2i\delta)\Gamma(1-2i\delta)} \cr
\times&\frac{\Gamma(1/2+\ts-2i\tomega-i\delta)\Gamma(1/2-\ts-2i\tomega-i\delta)}
{\Gamma(1/2+\ts-2i\tomega+i\delta)\Gamma(1/2-\ts-2i\tomega+i\delta)}\cr
&=(-2i\tomega\sigma)^{2i\delta}
\frac{\Gamma(1/2+2i\tomega+i\delta-4i\tau/\sigma)}{\Gamma(1/2+2i\tomega-i\delta-4i\tau/\sigma)}.
\label{VIE-1}
\end{align}
Equation (\ref{VIE-1}) determines the QNM frequency $\tomega$. 

We have assumed that $\tomega\rightarrow m/2$ in the limit of $a\rightarrow M$ 
or equivalently $\epsilon\rightarrow0_\pm$. It is known that $\delta$ is real and positive in this limit, i.e., 
extremal black hole case, for $|\ts|=2$ and $l=|m|\geq2$ \cite{Teukolsky1974,Starobinski1974}, 
and we focus on such cases. Then, the left hand side of 
Es.~(\ref{VIE-1}) will have a finite limit for $\epsilon\rightarrow 0_\pm$. We write it in the form, 
${\rm L.H.S.}=q e^{i\chi}$.
In order that the right hand side has also a finite limit, 
$\tomega\propto \sigma^{-1}$ or $\tau/\sigma\rightarrow \infty$ should hold in this limit. 
Here note that $\sigma\rightarrow0$ 
in the limit of $\epsilon\rightarrow0_\pm$. This fact implies that the former is inconsistent with 
our assumption, and $\tau/\sigma$ should diverge in the limit of $\epsilon\rightarrow 0_\pm$, i.e., 
$p$ in Eq.~(\ref{assumption}) should satisfy $p<1/2$. 
Then, through completely the same arguments as that in Ref.~\cite{Cardoso2004}, we have
\begin{align}
\tomega_{\rm R}&\simeq\frac{m}{2}-\frac{1}{4m}e^{(\chi-2k\pi)/2\delta}\cos\zeta \cr
\tomega_{\rm I}&\simeq-\frac{1}{4m}e^{(\chi-2k\pi)/2\delta}\sin\zeta, \nonumber
\end{align}
for both the black hole and the superspinar, where $k$ is an integer number consistent with 
$\tomega_I={\cal O}(|\epsilon|^p)$.
The estimate of $\zeta$ by Sasaki and Nakamura \cite{Sasaki1990} is available not only for the black hole but also for 
the superspinar, in case of $\tomega\simeq m/2$; $0<\zeta<2$. 
Thus the imaginary part of the QNM frequency $\tomega_{\rm I}$ is negative.
This result implies that, under the condition (\ref{BC-near-1}), both the black hole and the superspinar 
are stable against the gravitational wave perturbations of $m=l$. We have found that there is at least one boundary condition 
under which the superspinar is stable against the gravitational perturbations of $m=l$, although the physical meaning of 
the boundary condition is unclear and needs to be further investigated.

\section{Further Consideration}

Substituting Eqs.~(\ref{calA}) and (\ref{calB}) into Eq.~(\ref{QNM-condition}), we obtain 
\begin{align}
&D\sigma^{\ts}\left[{\cal F}(-\ts,\delta,-\tau/\sigma,-\tomega)+{\cal F}(-\ts,-\delta,-\tau/\sigma,-\tomega)\right] \nonumber\\
&=-C\sigma^{4i\tau\over\sigma}\left[{\cal F}(\ts,\delta,\tau/\sigma,\tomega)+{\cal F}(\ts,-\delta,\tau/\sigma,\tomega)\right], \label{general}
\end{align}
where,
\begin{align}
&{\cal F}(\ts,\delta,\tau/\sigma,\tomega) = 
(-2i\tomega)^{-i\delta}\Gamma(2i\delta)\Gamma(1+2i\delta) \nonumber\\
&\times\Gamma(1-\ts+4i\tau/\sigma) \bigl[\Gamma(1/2-\ts-2i\tomega+i\delta) \nonumber\\
&\times\Gamma(1/2-\ts+2i\tomega+i\delta)\Gamma(1/2-2i\tomega+i\delta+4i\tau/\sigma)\bigr]^{-1}. \nonumber
\end{align}

As mentioned below Eq.~(\ref{QNM-condition}), in contrast to the black hole, 
there are no conditions to determine $C$ and $D$ in the superspinar case, since 
there is no singular point in Eq.~(\ref{near-2}) on the real axis of $r$.  
This fact implies that there is no physical requirement that determines 
$\tomega$ in the superspinar case. Hence we replace the question from usual one, i.e., ``Which sign does $\tomega_{\rm I}$ have?" 
Since we do not know the physical nature of the superspinar, we would like 
to ask, {\it ``Are there boundary conditions under which the superspinar is stable?"}
If such boundary conditions exist, the stable superspinar will have a physical nature 
which leads to one of such boundary conditions. 

The answer to this new question is ``Yes", since we may regard $\tomega$ as an input parameter 
and Eq.~(\ref{general}) as an equation to determine the ratio between $D$ and $C$: We may assume
$\tomega=m/2+i\tomega_{\rm I}$ with $\tomega_{\rm I}={\cal O}(|\epsilon|^p)<0$.
Since $\tomega_{\rm I}$ is arbitrary as long as it is negative and ${\cal O}(|\epsilon|^p)$, we therefore have an infinite number 
of boundary conditions under which the superspinar is stable. Once the ratio between $C$ and $D$ 
is determined through Eq.~(\ref{general}), we have the ratio between $A$ and $B$ 
through the matching condition (\ref{matching}). 
As a result, we have a damping solution of quasi-normal mode  
and can find the boundary condition at, for example, $y=0$ by using this solution. 

The present analysis is restricted to the modes of $m=l$ for the near-extremal case. 
However, the situation is the same for general case, since there is no singular point of Eq. (\ref{tR-eq}) on the real axis of $r$ in the superspinar case. 
Unless we have some physical reason to impose the boundary condition at 
some $r$, we cannot say anything specific about the stability of the superspinar. 

The ergo-region instability is well known for a rapidly rotating compact object 
in which there is neither source nor absorber of the energy flux of perturbations \cite{Friedman1978,Comins1978}.  
Hence, it has been thought that a stable compact object with an ergo-region can only be a black hole. 
However, the present analysis suggests that this is not necessarily true, 
since we have not excluded a possibility that there is a source or an absorber of the energy flux ``inside the superspinar". 
There are infinite kinds of stable compact object with ergo-regions, although we do not know whether they are composed of the physically reasonable matter.

\bigskip

{\it Acknowledgments}: KN and PSJ are grateful to V. Cardoso for very useful comments. 
KN thanks colleagues at the elementary particle physics 
and gravity group in Osaka City University 
and participants to Goshiki-hama workshop held in 2016 for useful discussions. 
This work was supported in part by JSPS KAKENHI Grant Numbers~JP25400265 (KN) and 
JP26400282 (TH). 
PSJ, JG and PK like to thank Osaka City University where part of the work was done.

\end{document}